\documentstyle[eqsecnum,aps,multicol]{revtex} 

\begin{document}
\draft
\preprint{}
\title{Velocity Statistics in the Two-Dimensional Granular Turbulence
}
\author{Masaharu Isobe
\footnote{Email address: isobe@nitech.ac.jp
}
}
\address{
Graduate School of Engineering, Nagoya Institute of Technology, 
Nagoya, 466-8555, Japan
}

\date{\today}
\maketitle
\begin{abstract}
We studied the macroscopic statistical properties on the freely evolving quasi-elastic hard disk (granular) system by performing a large-scale (up to a few million particles) event-driven molecular dynamics systematically and found that remarkably analogous to an enstrophy cascade process in the decaying two-dimensional fluid turbulence.
There are four typical stages in the freely evolving inelastic hard disk system, which are homogeneous, shearing (vortex), clustering and final state.
In the shearing stage, the self-organized macroscopic coherent vortices become dominant.
In the clustering stage, the energy spectra are close to the expectation of Kraichnan-Batchelor theory and the squared two-particle separation strictly obeys Richardson law.
\end{abstract}

\begin{multicols}{2}

%%%%% Introduction
The dynamics of granular materials becomes one of the most important topics in the studies of nonlinear, dissipative, and non-equilibrium statistical physics\cite{kadanoff_1999}.
Granular media are collections of macroscopic particles with rough surfaces, dissipative, and frictional interactions.
The granular systems require an energy source in order to be in a steady state and the external gravitational force much affect their dynamics.

%%% IHS model
To focus on the dissipative features, smooth inelastic hard sphere (IHS) model are often used as an ideal model.
The freely cooling granular fluid has been studied as an ideal dissipative particle system in the absence of external force.
Since the system is only composed of an inelastic hard spheres and no relevant energy scale exists, the restitution coefficient between collision particles is the only parameter in terms of non-equilibrium.
The assumption of an inelastic hard sphere potential is also employed in kinetic theory, which facilitates comparisons between theory and simulation.
In order to construct the theory of the macroscopic phenomenology in non-equilibrium dissipative particle system, an IHS model is the most promising as a microscopic model.
A linear stability analysis of hydrodynamic equations for IHS model has revealed that the initial spatially homogeneous cooling state is unstable to the formation of vortices and clusters.
The shearing (vortex) and cluster instabilities were theoretically predicted and were tested by molecular dynamics (MD) simulations~\cite{goldhirsch_1993}.

%%% heat bath
Since the total energy is monotonically decreasing in the freely evolving process, a steady state in terms of energy fluctuation can be realized by scaling the velocity of the entire particle to the total energy remaining constant \cite{komatsu_2000}. 
Here, we introduce the new-scaled time $t_s$ which is described by

\begin{equation}
t_s = \int_0^t \frac{dt}{\beta(t)}, \quad \beta(t) = \sqrt{T(0)/T(t)},
\end{equation}

\noindent
where $T(t)$ is the average kinetic energy per particle as a function of usual time $t$.
The great advantage of this scaling operation is that the trajectories of particles don't change compared with the non-scaled case in case of hard sphere system.
Therefore, one simply replaces the usual time $t$ with the new-scaled time $t_s$. 
This operation is the same as the well-known velocity scaling method~\cite{woodcock_1971} in the usual MD simulation, in which the velocities of all particle ${\bf v}_i(t)$ are scaled following each collision by the factor $\beta(t)$ (i.e. $\beta(t){\bf v}_i(t)$) and the total energy is kept fixed strictly all over the time.
The 2D turbulence in nature is a large-scale fluid motion in the atmosphere or ocean dynamics on earth.
The most remarkable feature of 2D turbulence is described by the enstrophy cascade dynamics, which is completely different from that of 3D turbulence represented by the K41 theory.
The existence of enstrophy cascade process was originally proposed by Kraichnan\cite{kraichnan_1967} and Batchelor\cite{batchelor_1969}.
The theory expected that the enstrophy injected at a prescribed scale is dissipated at smaller scales, undergoing a cascading process at a constant enstrophy transfer rate; this led to predicting a $k^{-3}$ spectrum for the energy, in a range of scales extending from the injection to the dissipative scale.
The granular kinetic energy spectrum in the connection with the fluid turbulence was firstly pointed out by Taguchi~\cite{taguchi_1993} in 2D granular vibrated beds.
He obtained the results of the $k^{-5/3}$ spectrum in his simulation with a few hundred particles.
Another important nature of 2D turbulence is the self-organized coherent vortices, which develop into larger ones through the merging process between vortices with same sign of circulation.

%%% scope of this paper
In this letter, we especially focus on the velocity statistics and statistical laws on the fluid turbulence.
To specify what is the universal characters in dissipative system both macroscopic equation and microscopic dissipative particle, we performed extensive event-driven molecular dynamics simulation systematically on a freely cooling process in 2D IHS model with velocity scaling thermostat.
We found a strong similarity between 2D IHS model and 2D NS fluid turbulence.

%%%%% Model & Numerical Setting 
The freely 2D IHS (granular) model is so simple that the system is completely characterized with only three dimensionless parameters: the restitution coefficient $r$, the total number of disks $N$, and the packing fraction $\nu$. 
The system size in the unit of disk diameter $d$ is $L/d=\sqrt{\pi N/\nu}/2$.
All the disks are identical, namely the system is monodispersed.
The collision is instantaneous and only binary collisions occur.
When two disks, $i$ and $j$, with respective velocities ${\bf v}_i$ and ${\bf v}_j$ collide, the velocities after the collision, ${\bf v}_i'$ and ${\bf v}_j'$, are given by

\begin{eqnarray}
{\bf v}_i' & = &
{\bf v}_i-\frac{1}{2}(1+r)[{\bf n}\cdot ({\bf v}_i-{\bf v}_j)]{\bf n}
\\
{\bf v}_j' & = &
{\bf v}_j+\frac{1}{2}(1+r)[{\bf n}\cdot ({\bf v}_i-{\bf v}_j)]{\bf n},
\end{eqnarray}
       
\noindent
where ${\bf n}$ is the unit vector parallel to the relative position of the two colliding disks in contact.
Our system consists of more than 250 thousands hard disks (up to a few million) placed in a square box with periodic boundaries without no external force.
To perform such a large-scale simulation, we implemented the simple and efficient event-driven algorithm, which can actually simulate more than a few million particles even in the personal computer \cite{isobe_1999}.
Initially, the system is prepared as the equilibrium state by the long enough preliminary run with the restitution coefficient $r=1$, in which the density is uniform and the disk velocities are Maxwell-Boltzmann distribution.
The packing fraction and the restitution coefficient ($\nu,r$) were varied from dilute to dense and within shearing regime, which is estimated by the criterion of McNamara and Young~\cite{mcnamara_1996}, respectively.
In case of $(N,\nu) =(1024,0.25)$, McNamara and Young have found the final states have three typical states, which are kinetic, shearing, and collapse regime.
However, the spatio-temporal structure of shearing regime have not known yet especially at the macroscopic level.

The criterion of kinetic-shearing boundary in Ref.~\cite{mcnamara_1996} is based on the results of Jenkins and Richman~\cite{jenkins_1985}, in which the high wave number cutoff for the unstable shear modes were derived.
On the contrary, the shearing-collapse boundary is estimated by 1D theoretical analyses for inelastic collapse, which are known as the phenomenon on the number of collisions during a finite time diverges.
By using the the regime criterion described by McNamara and Young~\cite{mcnamara_1996}, one can find both regime boundaries become close to the unity in the thermodynamic limit.
These theoretical expectation indicate that the system always unstable even in the quasi-elastic limit.
This fact implies the important conjecture discussed later when we consider a large-scale IHS model as the macroscopic fluid model.

%%%%% Results %%%%%%
%%% Figure 1: Relaxation process
%%% Figure 2: Typical snapshots for stage 2 & stage 3
In the large-scale simulations, the restitution coefficients within shearing regime become quasi-elastic ($r\sim1$).
All our simulations by changing various parameters within the shearing regime, the system evolves to the final steady state after several stages.
As described in McNamara and Young~\cite{mcnamara_1996}, we can calculate the packing fraction $\nu({\bf x})$, velocity ${\bf u}({\bf x})=(u_x({\bf x}),u_y({\bf x}))$, and temperature $T({\bf x})$ at any point ${\bf x}$.
Figure 1 represents typical evolving process for four normalized properties as a function of new-scaled time $t_s$ in 2D IHS model.
During relaxation process, four stages can be distinguished.
After the homogeneous cooling state (HCS) continues for a certain time from initial thermal equilibrium state(first stage), the short-range velocity correlation for each time (i.e. both pre- and postcollisional velocity correlation) within the distance $\sum_{\bf x} {\bf u}({\bf x})\cdot {\bf u}({\bf x'})$ sharply deviate from zero (solid line in Fig.1), where ${\bf x'}$ is location around ${\bf x}$ with a distance of disk diameter $d$ (second stage).
Note that there are several works on the existence for the short-range velocity correlations even in HCS~\cite{soto_2001,nakanishi_2003}.
Our simulation also seems to show that velocity correlation 'gradually' increase from the beginning of simulation in first stage.
Therefore, it might to be difficult to determine the exact time between first and second stage.
In this stage, coherent vortices self-organize and the coherent vortices develop into larger ones through the mutual confluence and the merging process among vortices with same sign of circulation (Fig.~2 (a)).
These self-organized vortices are found firstly by McWilliams~\cite{mcwilliams_1984} in the direct numerical simulation (DNS) of 2D NS fluid turbulence. 
In the third stage, the density fluctuation $\nu_{rms}$ (dotted line in Fig.1), which is calculated the square root of the space average $(\nu({\bf x})-\nu)^2$, exhibit instability compressive flow (Fig.~2 (b)).
Finally, steady state is realized (fourth stage).
In the final steady states, the spatial correlation of inelastic hard 
disks, which gradually increases from the beginning of simulation, might be reached beyond the system size and begin to interfere with each other though the periodic boundary condition. 
Our simulations, in the shearing regime, there is no sign for the inelastic hard disks assembling to one cluster during the simulation time. This is because the shear mode expanded to the whole system might be stable though the periodic boundary condition.
Therefore, we call them `final steady states'.
We found there are four characteristic final steady states, which are shear (laminar, oscillating, and turbulent) and vortex (one pair of vortices with opposite sign of circulation) flows, by changing both packing fraction and the restitution coefficient within shearing regime, systematically.
Vortex flows of final pattern are also observed in the DNS simulation for 2D incompressible turbulence with the periodic boundary condition.

The velocity anisotropy $A=\sum (u_x({\bf x})^2-u_y({\bf x})^2)/(u_x({\bf x})^2+u_y({\bf x})^2)$, in which one can distinguish the final state as a shear ($A=-1$ or $1$) or a vortex ($A=0$) flows, and the enstrophy $Z = \sum |\omega({\bf x})|^2$, where $\omega({\bf x}) = rot\ {\bf u}({\bf x})$, are also plotted by dot dashed line and dashed line in Fig. 1, respectively.
We confirmed that the total vorticity $\sum_{\bf x} \omega ({\bf x})$ is zero throughout the simulation.

%%% Figure 3: Energy spectrum on 2d turbulence
The 2D fluid turbulence has different characters on the statistical law between forced and freely decaying case.
However, in the granular case, no systematic consideration seems to exist yet.
In the previous studies related to energy spectrum in granular material, energy injection (thermostat) are driven by the vibration cycle~\cite{taguchi_1993} and the periodical-stochastic thermostat~\cite{peng_1998}.
These energy injections resemble in the studies of forced fluid turbulence.
On the other hand, velocity scaling thermostat~\cite{komatsu_2000}, in which the system is driven continuously, is thought as corresponding to a freely decaying case, because the statistics itself don't change by introducing the new-scaled time $t_s$.
Actually, we found the energy spectra (power spectra of velocity field) in 2D IHS model with the velocity scaling thermostat are close to the expectation of Kraichnan-Batchelor theory ($E(k)\sim k^{-3}$) after the third stage (Fig. 3).
Therefore, our simulations show the enstrophy cascade, as is expected by the theory for freely decaying 2D fluid turbulence.
We have confirmed this power law by changing several different system size.
In Fig. 3, we can estimate the characteristic spatial scale ($k_d\sim 0.3$) for minimal dissipative domain (such as Kolmogorov scale in the fluid turbulence), which is composed of about a thousand disks.

%%% Figure 4: Enstrophy power-low & Richardson law
The first quantitative phenomenological observation in developed turbulence was shown by Richardson~\cite{richardson_1926}, in which the two fluid particle separation $R=|{\bf r}_i-{\bf r}_j|$ obeys power law  ($\langle R^2\rangle \sim t^3$).
We also show that the time ($t_s$) dependence of the space-averaged squared two-disk separation in 2D IHS model strictly obeys Richardson dispersion law in the third stage (Fig. 4).
In the inset of Fig. 4, the enstrophy evolution is also plotted in terms of the new-scaled time $t_s$.
The enstrophy seems to decay power-law behavior in the second stage, in which the coherent vortices self-organize.
However, since the second stage itself is relatively short, this behavior needs further confirmation.
%%% Flatness of vorticity
As the intermittency of vorticity, McWilliams found that the probability distribution function of vorticity significantly deviate from Gaussian~\cite{mcwilliams_1984}.
By calculating flatness of vorticity ($f_\omega = \langle \omega^4 \rangle /\langle \omega^2 \rangle ^2$) in 2D IHS model, our simulations also show the deviation from Gaussian after third stage.

%%% comments for future papers
How should we understand the obtained results?
The different points between fluid turbulence and granular turbulence are compressibility, the origin of dissipation (i.e. viscosity and inelasticity between particles) and the ratio of particle size and system size.
Is the granular turbulence close to fluid turbulence in the thermodynamic limit?
Let us assume the extreme condition that is dense, thermodynamic and elastic limit.
In this situation, a little amount of dissipation in the system always results in unstable state.
We also obtained the fact that the velocity correlation length (Kolmogorov scale) becomes larger in dense (i.e., quasi-incompressible) system, but less than system size when we consider the thermodynamic limit.
Therefore, this extreme condition seems to really correspond to NS fluid turbulence.

%%%%% Summaries & Future Works %%%%%
In this letter, we showed the remarkable similar aspects on the statistical law of vorticity between 2D IHS (granular) turbulence and 2D NS fluid turbulence.
These results were obtained by only solving a simple Newton's equation system for inelastic hard disks in terms of event-driven scheme.
From the microscopic dynamics of inelastic hard disk to the macroscopic fluid, it is important to study the origins of statistical law for turbulence at the microscopic level, but there are very few studies so far from this point of view.
The discussion for three limits (dense, thermodynamic, and elastic) might make a connection between 2D granular turbulence and 2D fluid turbulence.

I would like to thank to Profs. Y.~Hiwatari, H.~Nakanishi, H.~Hayakawa, and S.~Nos\'e for making helpful comments.
I also acknowledge helpful discussion with Prof. T.~Gotoh and Dr. T.~Watanabe on the two dimensional fluid turbulence.
This work was partially supported by the Ministry of Education, Science, Sports and Culture, Grant-in-Aid for Scientific Research (C), 15560042, 2003.

\begin{figure}
\caption{
The time evolution of various properties from a equilibrium, which are density fluctuation (dotted line), velocity correlation within a short distance (solid line), anisotropy of total velocity (dot dashed line), and enstrophy (dashed line), respectively.
The $O(t_s)$ indicated four normalized properties as a function of $t_s$.
The parameters are fixed at ($r, N, \nu$)=($0.99109, 512^2, 0.60$) during the simulation. 
An inset in the below-left-hand corner shows the early stage of time evolution when shearing and clustering instability appears.
}
\label{fig:1}
\end{figure}

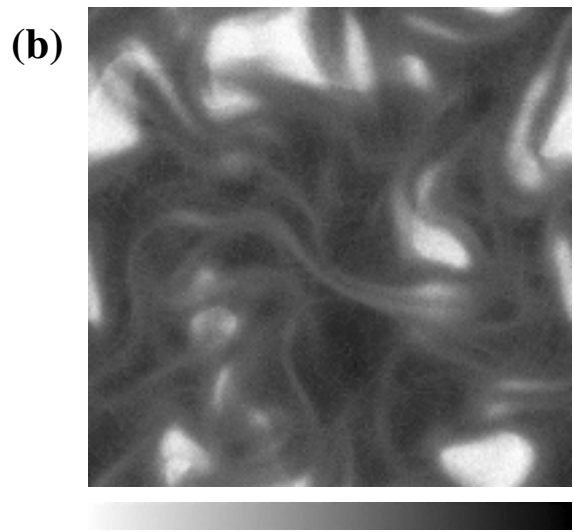
\begin{figure}
\caption{
The typical snapshots for coherent vortices and turbulent clustering patterns are shown.
(a) The absolute vorticity field after 1200 collisions per particle with ($r, N, \nu$)=($0.99452, 640000, 0.65$).
The self-organized coherent vortices in the vorticity field grow spontaneously from the initial equilibrium state.
(b) The density field after 3590 collisions per particle in inelastic hard disk system with 
($r, N, \nu$)=($0.99725, 2560000, 0.65$). 
The turbulent compressive flow appears in density field. 
}
\label{fig:2}
\end{figure}

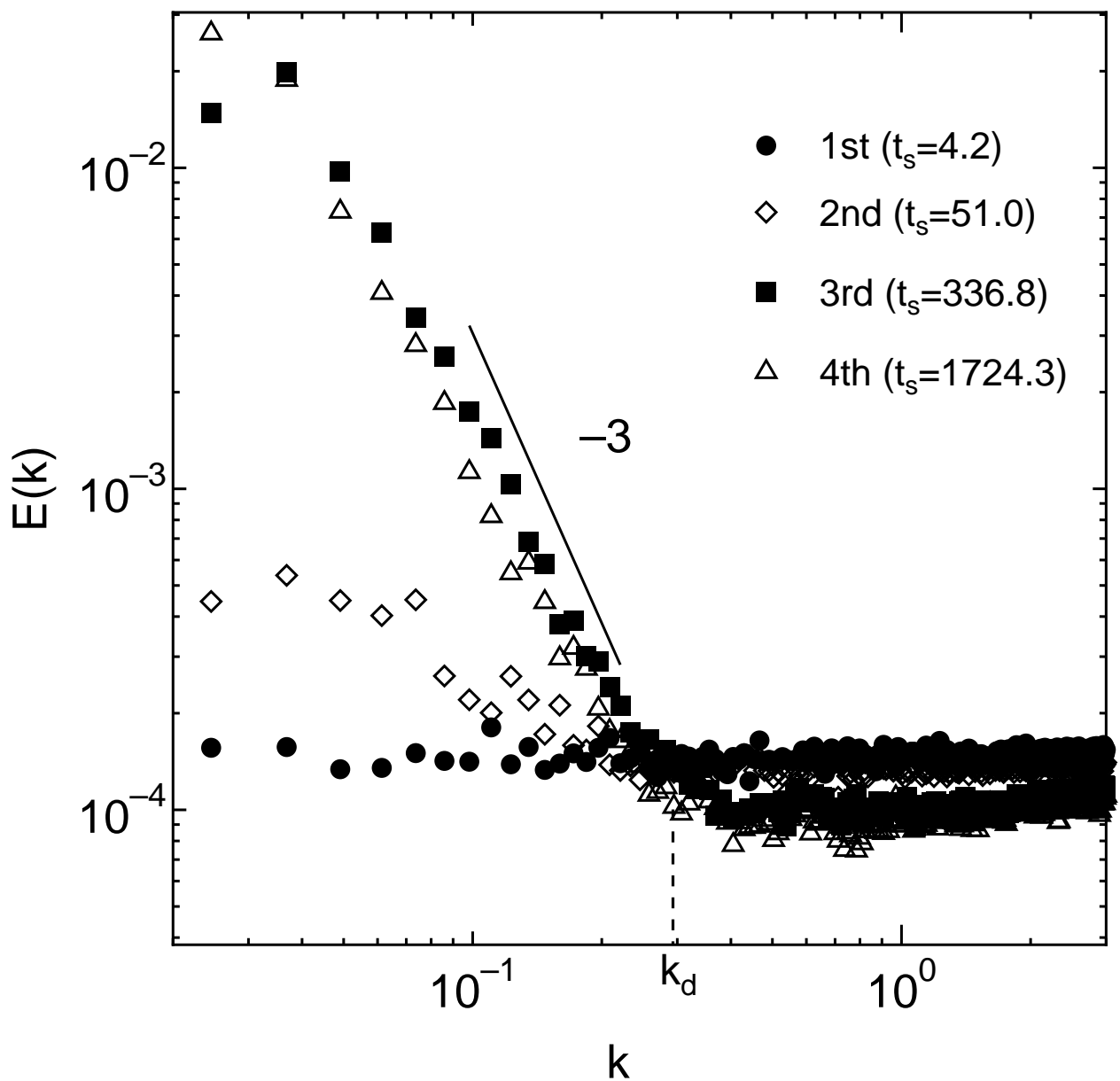
\begin{figure}
\caption{
Energy spectra of the velocity field are plotted for each stage.
The parameters are at ($r, N, \nu$)=($0.99109, 512^2, 0.60$). 
}
\label{fig:3}
\end{figure}

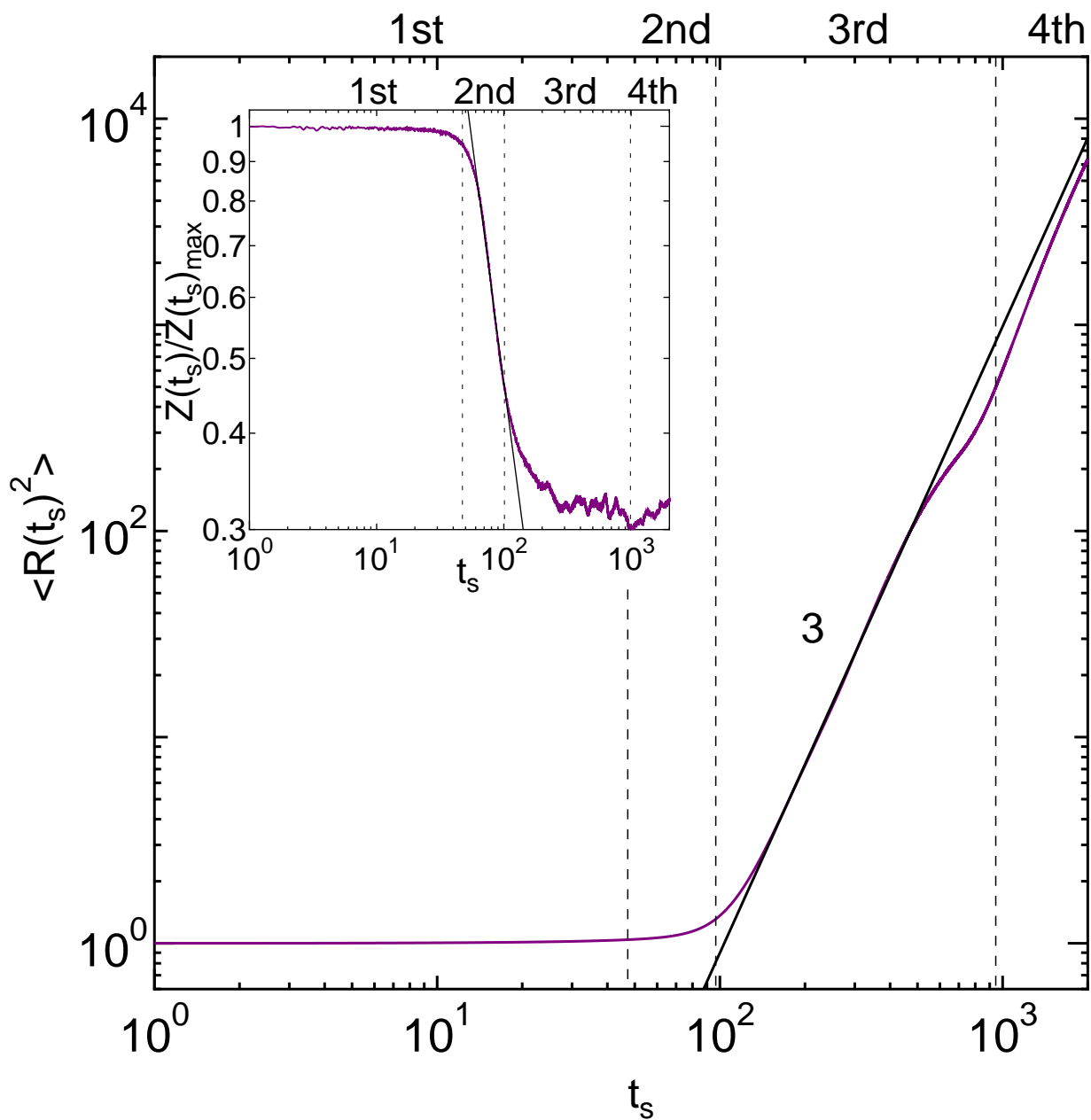
\begin{figure}
\caption{
The evolving squared two-particle separation in terms of new-scaled time $t_s$ is plotted.
An inset in the upper-left hand corner shows the time dependence of enstrophy decay.
The parameters are at ($r, N, \nu$)=($0.99109, 512^2, 0.60$). 
}        
\label{fig:4}
\end{figure}

\end{multicols}

\begin{references}

\bibitem{kadanoff_1999} L. P. Kadanoff,
 Rev. Mod. Phys. {\bf 71}, 435 (1999).

\bibitem{goldhirsch_1993} I. Goldhirsch, M. -L. Tan, and G. Zanetti,
 J. Sci. Comp. {\bf 8}, 1 (1993); I. Goldhirsch and G. Zanetti,
 Phys. Rev. Lett. {\bf 70}, 1619 (1993).

\bibitem{komatsu_2000} T. S. Komatsu,
 J. Phys. Soc. Jpn. {\bf 69}, 5 (2000); 
R. Soto, M. Mareschal, and M. M. Monsour,
 Phys. Rev. E {\bf 62}, 3836 (2000).

\bibitem{woodcock_1971} L. V. Woodcock,
 Chem. Phys. Lett. {\bf 10}, 257 (1971).

%\bibitem{hoover_1982} W. G. Hoover, A. J. C. Ladd, and B. Moran,
% Phys. Rev. Lett. {\bf 48}, 1818 (1982).

%\bibitem{evans_1983} D. J. Evans,
% J. Chem. Phys. {\bf 78}, 3297 (1983).

%\bibitem{nose_1990} S. Nos\'e,
% Prog. Theor. Phys. Suppl. {\bf 103}, 1 (1991).

%\bibitem{bizon_1999} C. Bizon, M. D. Shattuck, J. B. Swift, and H. L. Swinney,
% Phys. Rev. E {\bf 60} 4340 (1999).

%\bibitem{montanero_2000} J. M. Montanero and A. Santos, 
% Granular Matter {\bf 2}, 53 (2000).

%\bibitem{ernst_2002} M. H. Ernst and R. Brito,
% Phys. Rev. E {\bf 65}, 040301 (2002).

%\bibitem{muller_1998} I. M\"uller and T. Ruggeri,
% {\it Rational Extended Thermodynamics}, (Springer, 1998)

\bibitem{kraichnan_1967} R. H. Kraichnan,
 Phys. Fluids {\bf 10}, 1417 (1967).

\bibitem{batchelor_1969} G. Batchelor,
 Phys. Fluids Suppl. II {\bf 12}, 233 (1969).

\bibitem{taguchi_1993} Y. -h. Taguchi,
 Europhys. Lett. {\bf 24}, 203 (1993).

\bibitem{isobe_1999} M. Isobe,
 Int. J. Mod. Phys. C {\bf 10}, 1281 (1999).

\bibitem{mcnamara_1996} S. McNamara and W. R. Young, 
 Phys. Rev. E {\bf 53}, 5089 (1996).

\bibitem{jenkins_1985} J. T. Jenkins adn M. W. Richman,
 Arch. Ration. Mech. Anal. {\bf 87}, 355 (1985).

\bibitem{soto_2001} R. Soto and M. Mareschal,
 Phys. Rev. E {\bf 63}, 041303 (2001);
R. Soto, J. Piasecki, and M. Mareschal,
 Phys. Rev. E {\bf 64}, 031306 (2001).

\bibitem{nakanishi_2003} H. Nakanishi, 
 Phys. Rev. E {\bf 67}, 010301(R) (2003);
 R. Kawahara and H. Nakanishi, preprint.

\bibitem{mcwilliams_1984} J. C. McWilliams,
 J. Fluid Mech. {\bf 146}, 21 (1984).

\bibitem{peng_1998} G. Peng and T. Ohta,
 Phys. Rev. E {\bf 58}, 4737 (1998).

\bibitem{richardson_1926} L. F. Richardson, 
 Proc. R. Soc. London Sect.A {\bf 110}, 709 (1926).

%\bibitem{rutgers_1998} M. A. Rutgers,
% Phys. Rev. Lett. {\bf 81}, 2244 (1998).

%\bibitem{alam_2001} M. Alam and C. M. Hrenya,
% Phys. Rev. E {\bf 63}, 061308 (2001).


\end{references}
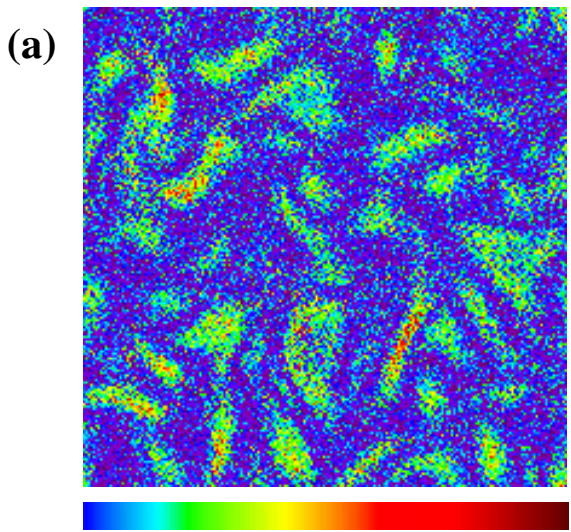
\end{document}